\documentclass[10pt]{article}
\usepackage{amsmath, amssymb}
\usepackage[OE]{expresss}

\usepackage{graphicx}

\newcommand\dd {\, \mathrm{d}}
\newcommand\ddd {\mathrm{d}}

\newcommand{\mat}{\mathbf}

\newcommand\degg{{}^\circ}

\newcommand\mm{\textrm{~mm}}

\begin{document}

\title{Designing double freeform surfaces 
\\for collimated beam shaping with optimal mass transportation 
and linear assignment problems}

\author{Leonid~L.~Doskolovich,\!\authormark{1,2,*} Dmitry~A.~Bykov,\!\authormark{1,2} Evgeniy~S.~Andreev,\!\authormark{1,2} Evgeni~A.~Bezus,\!\authormark{1,2} and Vladimir~Oliker\authormark{3}}
\address{\authormark{1}Image Processing Systems Institute --- Branch of the Federal Scientific Research Centre ``Crystallography and Photonics'' of Russian Academy of Sciences, 151 Molodogvardeyskaya st., Samara 443001, Russia\\
\authormark{2}Samara National Research University, 34 Moskovskoye shosse, Samara 443086, Russia\\
\authormark{3}Emory University, Atlanta, GA 30322, USA}
\email{\authormark{*}leonid@smr.ru}

\begin{abstract}We propose a method for designing refractive optical elements for collimated beam shaping. In this method, the problem of finding a ray mapping is formulated as a linear assignment problem, which is a discrete version of the corresponding mass transportation problem. A method for reconstructing optical surfaces from a computed discrete ray mapping is proposed. The method is suitable for designing continuous piecewise-smooth optical surfaces. The design of refractive optical elements transforming beams with circular cross-section to variously shaped (rectangular, triangular, and cross-shaped) beams with plane wavefront is discussed. The presented numerical simulation results confirm high efficiency of the designed optical elements.
\end{abstract}


\section{Introduction}
The problem of collimated beam shaping consists in the design of a refractive optical element transforming a given incident beam with a plane wavefront into an output beam with a prescribed irradiance distribution and a plane wavefront. In the general case, an optical element that performs this transformation has two working surfaces. As a rule, the problem under study is addressed within geometrical optics and has a wide range of applications, such as illumination, laser beam processing and optical lithography. The problem is easy to solve only in the axisymmetric case. In this case, the problem is reduced to a single spatial dimension, enabling the optical surfaces to be calculated by solving ordinary differential equations~\cite{1,2,3,4,5}. In the general two-dimensional (2D) case, the design of an optical element is essentially more challenging.

In Refs.~\cite{6,7,8}, the problem of collimated beam shaping was reduced to finding a solution to an elliptic nonlinear partial differential equation (PDE)~\cite{6,7,8}. The PDE is solved using finite-difference methods, in which the derivatives are replaced by their finite-difference approximations. Thus, the solution of the PDE is reduced to the solution of a system of nonlinear equations with respect to the values of the sought-for function, which describes the first optical surface and is defined on a certain grid. The resulting system of equations is solved using Newton's method. The second optical surface is then reconstructed analytically from the first surface by applying the requirement of a constant optical path length. General shortcomings of this approach include high computational complexity of solving a system of nonlinear equations as well as the choice of the initial approximation to the solution of the system and the formulation of the boundary conditions.

The considered problem can also be addressed by using another, mathematically less rigorous approach based on the so-called ray-mapping methods~\cite{9,10,11}. In this case, the ray mapping, i.e. the mapping between the coordinates of the rays on the wavefronts of the incident and the output beams is found by solving a standard Monge--Amp{\`e}re (MA) equation. The MA equation is solved either using the above-mentioned finite-difference method based on reducing the task to the solution of a system of nonlinear equations, or using an explicit finite-difference method based on the replacement of the original equation by the non-stationary parabolic Monge--Amp{\`e}re equation~\cite{10,11}. The optical surfaces are calculated on the basis of the computed ray mapping using numerical integration. In addition to having high computational complexity, the described ray-mapping methods can offer only an approximate solution because the standard MA equation is valid only in the paraxial approximation~\cite{8,12}.

In the basic theoretical works~\cite{13,14}, the problem of collimated beam shaping was formulated as a Monge--Kantorovich mass transportation problem, and the cost functions for the design of a refractive optical element~\cite{13} and of a two-reflector system~\cite{14} were derived. It is important to note that the cost function for the refractive optical element is non-quadratic~\cite{13}, meaning that in this case the ray mapping is not described by a standard MA equation. In a discrete version, the mass transportation problem can be formulated as a linear assignment problem (LAP)~\cite{15}. Such an approach is utilized in a recent work by the present authors~\cite{12} for solving an inverse problem of calculating the eikonal of a light field.

In this work, we for the first time apply this LAP-based approach to a problem of collimated beam shaping. 
Due to the fact that the element under study has two working optical surfaces, the application of the LAP-based method for the solution of this problem has a number of specific features and differs from~\cite{12}. 
As examples, we design refractive optical elements transforming a plane circular beam into plane beams with uniform irradiance distribution in a rectangular and a triangular domains in a nonparaxial case. These examples demonstrate high efficiency of the proposed LAP-based approach: the computation time for square grids of $143 \times 143$ points on a standard PC is of about 3~minutes. In contrast, the calculation of such a mapping using finite-difference methods for the solution of a PDE of Monge--Amp{\`e}re type requires solving a system of $143^2=20449$ nonlinear equations and is by several orders of magnitude slower~\cite{16}. Besides, the computation of the ray mapping by solving a mass transportation problem makes it possible to design elements with piecewise-smooth continuous optical surfaces, as distinct from smooth optical surfaces resulting from the solution of an elliptic nonlinear PDE~\cite{6,7,8,9,10,11}. It widens the range of the problems that can be solved, in particular, enabling the generation of uniform irradiance distributions in complex domains with non-smooth boundaries. As a demonstration of such a capability, we design an element with a piecewise-smooth optical surface transforming a circular beam into a cross-shaped beam.

\section{Formulation of the problem}

\begin{figure}
	\centering
		\includegraphics{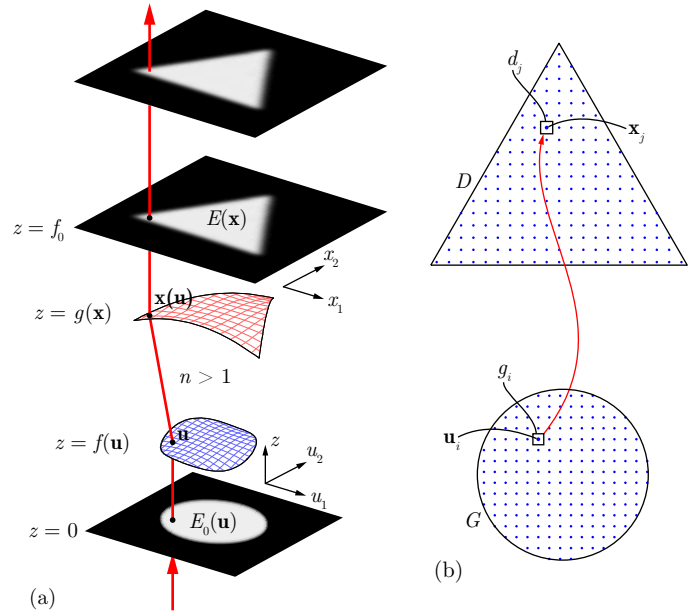}
	\caption{\label{fig1}(a)~Geometry of the problem of the design of an optical element. (b)~ Approximation of a circular domain $G$ and a triangular domain $D$ with equal-flux cells. Dots correspond to the centers of the cells.}
\end{figure}

Consider an incident beam with a plane wavefront perpendicular to the $z$ axis and with the irradiance distribution $E_0 (\mat{u}),\, \, \mat{u}\in G$, where $\mat{u}=(u_1 ,u_2 )$ are the Cartesian coordinates in the plane $z=0$ (Fig.~\ref{fig1}a).
The beam is being transformed by a refractive optical element with the refractive index $n > 1$, which is surrounded by a medium with the refractive index $n_0 = 1$. 
The element has two optical surfaces $f$ and $g$. The problem to be solved consists in the design of the optical surfaces generating an output beam with a plane wavefront perpendicular to the $z$ axis and a prescribed irradiance distribution $E(\mat{x}),\, \, \mat{x}\in D$, where $\mat{x} = (x_1, x_2)$ are the Cartesian coordinates in a certain plane $z = f_0$ located after the optical element. Note that the function $E(\mat{x})$  should satisfy the following normalization condition:
\begin{equation}
\label{GrindEQ__1_}
\int_{G}E_{0} (\mat{u})\dd\mat{u} =
 \int_{D}E(\mat{x})\dd\mat{x}, 
\end{equation}
where the integrals with respect to $\ddd\mat{u} \left(\ddd\mat{x}\right)$ denote double integrals with respect to $\ddd u_1 \ddd u_2 \left(\ddd x_1 \ddd x_2\right)$. 
Besides, since the output beam has a plane wavefront, the irradiance $E(\mat{x})$ will remain unchanged in any plane after the optical element (Fig.~\ref{fig1}a).
Moreover, since both the input and output wavefronts are plane, an optical path $L$ will be the same for any ray propagating from the plane $z=0$ to the plane $z=f_0$. It is convenient to define the value of $L$ through the optical element thickness $h_0$ at a certain point in the form $L = (n-1)h_0 + f_0$. 
The quantity $h_0$ can be treated as a parameter of the problem.

Let $\mat{x}=T(\mat{u}) = \left(x_1(u_1, u_2), x_2(u_1, u_2 )\right)$ denote a ray mapping that represents the Cartesian coordinates $\mat{x}$ of a ray in the output plane $z = f_0$ through the coordinates of the ray $\mat{u}$ in the plane $z=0$ (Fig.~\ref{fig1}a). According to the energy conservation law along the ray tubes, the mapping $T:G\to D$ conserves the light flux. The energy conservation law can be written in the integral form~\cite{12, 13}:
\begin{equation}
\label{GrindEQ__2_} 
\int_{\omega} E_0(\mat{u})\dd\mat{u}
=\int_{T(\omega)} E(\mat{x})\dd\mat{x},
\end{equation} 
where $\omega$ is any measurable subset of the domain $G$.

Note that for a ray to remain parallel to the $z$ axis after the refraction by the optical surfaces described by the functions $z=f(\mat{u})$ and $z = g(\mat{x})$, the normal vectors to the surfaces calculated at the points $\mat{u}$ and $\mat{x}(\mat{u}) = T(\mat{u})$ have to be parallel. In this case, the optical surfaces can be locally considered as a plane-parallel plate, which shifts a ray without changing its propagation direction. Using the vector form of the refraction law, the partial derivatives of the functions $z=f(\mat{u})$ and $z = g(\mat{x})$ can be easily derived in the form~\cite{13}
\begin{equation}
	\label{GrindEQ__3_}
	\begin{aligned}
	\frac{\partial f(\mat{u})}{\partial u_{i} } 
	&= -n\frac{x_i (\mat{u})-u_i }
	{\sqrt{(n-1)^2 h_{0}^2 -\left(n^2 -1\right)\left|\mat{x}(\mat{u})-\mat{u}\right|^2 } } ,\, \, i=1,2, \\ 
	\frac{\partial g(\mat{x})}{\partial x_{i} } 
	&= -n\frac{x_i -u_i (\mat{x})}
	{\sqrt{\left(n-1\right)^2 h_0^2 -\left(n^2 -1\right)\left|\mat{x}-\mat{u}(\mat{x})\right|^2 } } ,\, \, i=1,2,
	\end{aligned}
\end{equation}
where $\mat{u}(\mat{x})=T^{-1} (\mat{x})=\left(u_1(x_1, x_2), u_2(x_1, x_2)\right)$ is the inverse ray mapping. 
According to Eq.~\eqref{GrindEQ__3_}, $\frac{\partial f(\mat{u})}{\partial u_i } = \frac{\partial g\left(\mat{x}(\mat{u})\right)}{\partial x_i} ,\, \, i=1,2$. 
Besides, Eqs.~\eqref{GrindEQ__3_} suggest that not any mapping $\mat{x}(\mat{u})$ $\left[\mat{u}(\mat{x})\right]$ that conserves the energy in the sense of relationship~\eqref{GrindEQ__2_} is a solution of the initial problem.
Indeed, for the reconstruction of the functions $z=f(\mat{u})$ and $z=g(\mat{x})$ from Eqs.~\eqref{GrindEQ__3_}, the partial derivatives have to satisfy the following conditions:
\begin{equation}
\label{GrindEQ__4_}
\frac{\partial ^2 f(\mat{u})}{\partial u_{1} \partial u_{2} } =\frac{\partial ^2 f(\mat{u})}{\partial u_{2} \partial u_{1} },\, \, \, \frac{\partial ^2 g(\mat{x})}{\partial x_{1} \partial x_{2} } =\frac{\partial ^2 g(\mat{x})}{\partial x_{2} \partial x_{1} }.
\end{equation} 

It has been shown~\cite{13} that a mapping $\mat{x}=T(\mat{u})$ that corresponds to the solution of the problem in question satisfies the integrability conditions in Eq.~\eqref{GrindEQ__4_} and provides an extremum of the following functional:
\begin{equation}
\label{GrindEQ__5_}
P(T)=\int_G E_0(\mat{u})\,C(\mat{u}-T(\mat{u}))\dd\mat{u},
\end{equation}
where
\begin{equation}
\label{GrindEQ__6_}
C(\mat{s})=
\begin{cases}
-\sqrt{\gamma^2 -|\mat s|^2 }, & |\mat{s}|<\gamma; \\
\infty, & |\mat{s}|\geq\gamma,
\end{cases}
\end{equation}
where $\gamma = (n-1) h_0 / \sqrt{n^2-1}$.
The functional~\eqref{GrindEQ__5_} is defined on energy-conserving mappings $T:G\to D$.
According to Eqs.~\eqref{GrindEQ__5_} and~\eqref{GrindEQ__6_}, the mappings can be computed by solving a Monge--Kantorovich mass transportation problem. As the masses, the initial irradiance distribution $E_0(\mat{u}), \mat{u}\in G$ and the sought-for irradiance distribution $E(\mat{x}), \mat{x}\in D$ are used.
Let us note that the cost function defined by Eq.~\eqref{GrindEQ__6_} differs from the one utilized in~\cite{13} by an additive constant that does not affect the extremum of the functional~\eqref{GrindEQ__5_}.

As mentioned above, the optical surfaces in the vicinity of the points  $\mat{u}_0 $ and $\mat{x}(\mat{u}_0)$ locally correspond to a plane-parallel plate that shifts the ray going from the point $\mat{u}_0$ to the point $\mat{x}(\mat{u}_0)$ in the output plane.
It can be easily shown that the cost function in Eq.~\eqref{GrindEQ__6_} turns to infinity when, at given parameters $h_0$ and $n$ it is impossible to shift the ray by the magnitude $|\mat{s}_0| = |\mat{u}_0 - T(\mat{u}_0)|$. If the cost function turns to infinity on a set with nonzero measure, there may be no solution to the problem. Note that with the increase in the parameters $h_0$ and $n$ the achievable value of the shift also increases.

One can show that a mapping $\mat{x} = T(\mat{u})$ that minimizes the functional~\eqref{GrindEQ__5_} corresponds to a Galilean configuration of rays. The maximum of the functional corresponds to a mapping with a Keplerian configuration of rays.

In the paraxial approximation (i.e. under the condition $\left|\mat{x}(\mat{u})-\mat{u}\right|^2 \ll h_0 $) $C(\mat{s})\approx -\gamma +\frac{|\mat{s}|^2}{2\gamma}$, and the mapping $\mat{x}(\mat{u})=T(\mat{u})$ is computed by solving the mass transportation problem with a quadratic cost function. In this case, should there exist a differentiable mapping $\mat{x}(\mat{u})$, its computation is reduced to finding a solution to a standard Monge--Amp{\`e}re equation~\cite{8,9,10,11}.

It is important to note that the direct determination of the ray mapping from the solution of a mass transportation problem in Eqs.~\eqref{GrindEQ__5_} and~\eqref{GrindEQ__6_} makes it possible to operate with mappings that are differentiable not everywhere, but almost everywhere. In this case, the optical surfaces [functions $z=f(\mat{u})$ and $z=g(\mat{x})$] are continuous and piecewise-smooth, as distinct from smooth surfaces obtained as the solution of elliptic PDEs~\cite{6,7,8}. This makes it possible to extend the range of problems that can be solved, in particular, to generate prescribed irradiance distributions on a zero-intensity background in domains with complex and non-smooth boundaries.

\section{Reduction to the linear assignment problem}

The discrete version of the mass transportation problem in Eqs.~\eqref{GrindEQ__5_} and~\eqref{GrindEQ__6_} can be formulated as a linear assignment problem (LAP)~\cite{15}. Indeed, let us assume that the domains $G$ and $D$ are divided into $N$ equal-flux cells (or approximated by $N$ cells), with the equality $\int_{g_i} E_0 (\mat{u})\dd\mat{u}=\int_{d_j} E(\mat{x})\dd\mat{x}=e$ being valid for any pair of cells $g_i \subset G$, $d_j \subset D$. 
For the sake of illustration, Fig.~\ref{fig1}b shows approximations of a circular domain $G$ and a triangular domain $D$ by grids of $N = 196$ square cells. While constructing these approximations, the grid steps were chosen so that $N$ points fall within the domains $G$ and $D$. 
This approach based on the choice of the rectangular grid steps is suitable for generating uniform irradiance distributions in the domains of interest. 
In the case of nonuniform distributions $E_0 (\mat{u})$ and $E(\mat{x})$, an equal-flux set of $N = M_1 M_2$ cells in the domain $G$ (or $D$) can be constructed using the following general approach. The domain $G$ can be divided by parallel lines $u_1 =u_{1,i} ,\, i=1, \ldots, M_1$ into $M_1$ equal-flux sub-domains. Afterwards, each sub-domain $G_i$ is further divided into $M_2$ sub-domains by parallel lines $u_2 = u_{2,j}, j=1, \ldots, M_2$.

For the grids (approximations) introduced in the domains $G$ and $D$, all energy-conserving mappings $G\to D$ can be described by permutations $(i_1 , \ldots ,i_N )$ of $N$ numbers, which determine to which cells $d_i$ of the domain $D$ the cells $g_1 , \ldots,g_N$ are mapped. 
In terms of the LAP, the mapping of the cell $g_i$ to the cell $d_j$ can be interpreted as the assignment of the $j$-th task to the $i$-th agent. 
According to Eq.~\eqref{GrindEQ__6_}, the cost of the tasks is described by the matrix
\begin{equation} 
\label{GrindEQ__7_}
(\mat{M})_{i,j} = C(\mat{u}_i -\mat{x}_j ),\;\; i,\,j = 1,\ldots,N,
\end{equation} 
where $\mat{u}_i$, $\mat{x}_j$ are the centers of the cells $g_i$ and $d_j$. Thus, the mapping $\mat{x}(\mat{u}_i)=\hat{T}(\mat{u}_i)$ can be found by solving the following assignment problem:
\begin{equation} 
\label{GrindEQ__8_} 
P_d \left(j_1 , \ldots, j_N \right) = \sum_i C(\mat{u}_i - \mat{x}_{j_i}) \to \min,
\end{equation}
where the minimum is sought over all permutations $(j_1 ,\ldots,j_N )$. For the solution of the LAP, efficient polynomial algorithms have been proposed, including Hungarian algorithm~\cite{15}, Jonker--Volgenant algorithm~\cite{29}, and auction algorithm~\cite{17}.

\section{Reconstruction of the optical surface}\label{sec4}
Due to the discrete character of the LAP in Eq.~\eqref{GrindEQ__8_}, the resulting discrete mapping $\mat{x}_i =\hat{T}(\mat{u}_i), \, i=1,\ldots,N$ describes the sought-for continuous mapping $\mat{x}(\mat{u})=T(\mat{u})$ with an error. Besides, if the domain $D$ has a non-smooth boundary, the mapping $T(\mat{u})$ may have discontinuities. In this case, the reconstruction of the optical surfaces by direct numerical integration of Eqs.~\eqref{GrindEQ__3_} may result in a significant error.
Because of this, in this section we propose a method for reconstructing optical surfaces from a computed discrete mapping. Within the proposed method, the second optical surface [function $g(\mat{x})$] is assumed to be smooth, while the first surface [function $f(\mat{u})$] is assumed to be continuous and piecewise-smooth. This case corresponds to the most challenging example of an optical element described in subsection~\ref{sec53} and designed to generate a cross-shaped beam.

\subsection{Reconstruction of the upper optical surface}

Let us first consider the calculation of the function $g(\mat{x})$, which describes the second (upper) optical surface of the element. Because the function $g(\mat{x})$ is assumed to be smooth, it can be approximated by two-dimensional B-splines~\cite{19}.

Assume that the domain $D$, where $g(\mat{x})$ is defined, is bounded by a rectangle $\bar{D}$ with the sides $w_1$ and $w_2$. Let $x_1 = w_1 \cdot \left[m/(N_1-1) -1/2\right],\, m=0,\ldots ,N_1 -1$ be a set of $N_1$ equidistant knots on the interval $[-w_1 /2, w_1 /2]$ of the $x_1$ axis. Using these knots, let us introduce basis splines (B-splines) of the order $q$, denoting them by $B_m (x_1), \, m=1,\ldots, N_1 + q - 2$~\cite{19}. In a similar way, let us introduce B-splines $P_n(x_2)$ for the second coordinate $x_2 \in \left[-w_2/2, w_2/2\right]$. The function $g(\mat{x}) = g(x_1 ,x_2 )$ can be represented using the introduced basis functions as
\begin{equation}
\label{GrindEQ__9_}
g(\mat{x}) = \sum_{m,n} p_{m,n} B_m(x_1) P_n(x_2), 
\end{equation} 
where $p_{m,n}$ are the expansion coefficients. Note that if the splines $B_m(x_1)$ and $P_n(x_2)$ have the same order and the same number of knots, Eq.~\eqref{GrindEQ__9_} contains $N = (N_1 +q-2)^2$ unknown coefficients $p_{m,n}$.

The coefficients $p_{m,n}$ are determined from the condition that the values of the derivatives of the spline $g(\mat{x})$ at the points $\mat{x}_i =\hat{T}(\mat{u}_i),\, i=1,\ldots,N$ constitute a least-squares approximation of the derivatives $\frac{\partial \hat{g}}{\partial x_k} \left(\mat{x}_i \right)$. The latter are derived from Eq.~\eqref{GrindEQ__3_} on the basis of a discrete mapping $\hat{T}$, i.e. by assuming $\mat{u}(\mat{x}_i)=\hat{T}^{-1} (\mat{x}_i)$ in Eq.~\eqref{GrindEQ__3_}. Thus, the coefficients $p_{m,n}$ can be obtained by solving the following minimization problem:
\begin{equation}
\label{GrindEQ__10_}
S (\mat{p}) = \sum_i \left[\frac{\partial \hat{g}}{\partial x_1 } (\mat{x}_i)-\frac{\partial g}{\partial x_1} (\mat{x}_i )\right]^2  +\sum _{i}\left[\frac{\partial \hat{g}}{\partial x_2 } (\mat{x}_i)-\frac{\partial g}{\partial x_2 } (\mat{x}_i )\right]^2  \to \min ,
\end{equation} 
where $\mat{p}$ is the vector composed of the spline coefficients $p_{m,n}$. It is easy to show that the solution of the problem~\eqref{GrindEQ__10_} is reduced to a standard least-squares method. The proposed approach, which consists in approximating derivatives of an optical surface, allows one to compensate for the errors resulting from computing the mapping by use of the `discrete' LAP-based approach~\cite{20, 21}.

Note that for a nonrectangular domain $D\subset \bar{D}$, the solution of the problem~\eqref{GrindEQ__10_} may be numerically unstable. In this case, one should apply a regularized least-squares method, in which the term $\lambda \sum_{m,n}p_{m, n}^2$, where $\lambda $ is a small number, is added to the minimized function $S(\mat{p})$.

\subsection{Reconstruction of the lower optical surface}

Next, let us consider the calculation of the function $f(\mat{u})$, which describes the first (lower) optical surface. As we mentioned earlier, the function $f(\mat{u})$ is assumed to be piecewise-smooth, which does not allow us to approximate it by splines. We propose to reconstruct the function $f(\mat{u})$ through the function $g(\mat{x})$ using the representation of the optical surface from the supporting quadric method (SQM)~\cite{22,23,24,25}. The SQM is intended for designing a piecewise-smooth (refracting or reflecting) optical surface to focus the incident beam into a given array of points. In this case, the optical surface is composed of continuously stitched lens segments.

To apply the SQM to our problem, we assume that the second optical surface is defined on a sufficiently fine grid and is described by a set of points $\mat{X}_i =\left(\mat{x}_i ,g(\mat{x}_i) \right),\, i = 1, \ldots, N_g $. Using the spline representation of the optical surface $g(\mat{x})$ [Eq.~\eqref{GrindEQ__10_}], the array of points $\mat{X}_i$ can be calculated on a uniform grid with an arbitrarily small step. In the vicinity of each point $\mat{u}\in G$, the function $f(\mat{u})$ corresponds to a lens segment that focuses the incident plane beam into one of the points $\mat{X}_i$ on the upper optical surface.

To derive the lens equation, let us use Fermat's principle (the condition of a constant optical path length from the points on the incident wavefront to the focal point):
\begin{equation}
\label{GrindEQ__11_}
\Phi_i (\mat{u})+n\sqrt{\left|\mat{x}_i -\mat{u}\right|^2 +\left(g\left(\mat{x}_i \right)-\Phi_i (\mat{u})\right)^2 } = L_i,\;\;  i=1, \ldots, N_g , 
\end{equation}
where the function $\Phi_i (\mat{u})$ describes the lens surface and $L_i = L-\left(f_0 -g(\mat{x}_i )\right) = (n-1) h_0 +g(\mat{x}_i)$ is the optical path length of the rays propagating from the plane $z=0$ to the focal point $\mat{X}_i =\left(\mat{x}_i, g(\mat{x}_i)\right)$.
Equation~\eqref{GrindEQ__11_} defines the lens surface $z=\Phi_{i}(\mat{u})$ in an implicit form.
Through simple transformations, the function $\Phi_i (\mat{u})$ can be derived from Eq.~\eqref{GrindEQ__11_} in the following explicit form:
\begin{equation}
\label{GrindEQ__12_}
\Phi_i (\mat{u})=g(\mat{x}_i)  -\frac{1}{n+1} \left(h_0 +n \sqrt{h_0^2 -\frac{n+1}{n-1} \cdot \left|\mat{x}_i -\mat{u}\right|^2 } \right). 
\end{equation}
It is easy to show that the lens surface is described by an ellipsoid of rotation with the rotation axis parallel to the $z$-axis, eccentricity $\varepsilon = 1/n$, focal parameter $p = (n-1) h_0 /n$, and one of the foci located at the point $\mat{X}_i$~\cite{26}. 

According to the SQM, the surface that focuses into the set of points $\mat{X}_i$ consists of a set of ellipsoid segments defined by Eq.~\eqref{GrindEQ__12_}. At each point $\mat{u} \in G$, the surface is calculated from the following expression~\cite{22,23,24,25}:
\begin{equation}
\label{GrindEQ__13_}
f(\mat{u})=\mathop{\min }\limits_{i\in \{1, \ldots, N_g\}} \Phi_i (\mat{u}).
\end{equation}
Thus, we propose to calculate the function $f(\mat{u})$ using Eqs.~\eqref{GrindEQ__12_}--\eqref{GrindEQ__13_} through the second optical surface defined on a set of points $\mat{X}_i =\left(\mat{x}_i ,g(\mat{x}_i)\right),\, i=1,\ldots,N_g $. Note that Eq.~\eqref{GrindEQ__13_} is also convenient for calculating the mapping $T(\mat{u})$ at arbitrary values of $\mat{u}$. Indeed, according to Eq.~\eqref{GrindEQ__13_}, we obtain
\begin{equation}
\label{GrindEQ__14_}
T(\mat{u})=\mat{x}_j ,\, \, j=\mathop{\arg \min }\limits_{i\in \{1, \ldots, N_g\}} \Phi_i (\mat{u}).
\end{equation}

\section{Design examples}

\subsection{Transformation of a circular beam to a rectangular beam}

As a first example, we consider the design of an optical element transforming a beam with circular cross-section and uniform irradiance $E_0 (\mat{u}) = 1/(\pi R^2 )$, $\mat{u} \in G=\left\{\left. \left(u_1 ,u_2 \right)\,\right|\,u_1^2 +u_2^2 \le R^2 \right\}$ to a beam with rectangular cross-section and uniform irradiance $E(\mat{x})=1/(w_1 w_2 )$, $\mat{x}\in D=\left\{\left. \left(x_1 ,x_2 \right)\,\right|\,|x_1|\le w_1 / 2,\, |x_2|\le w_2 / 2\right\}$. 
The wavefronts of the incident and output beams are plane and perpendicular to the $z$ axis. 
The element was designed for the following parameters: the incident beam radius $R = 1\mm$, the sizes of the output rectangular beam $w_1 = 5\mm$ and $w_2 = 2.5\mm$, the optical element thickness $h_0 = 5\mm$ at $\mat{u} = (0,0)$, and the refractive index of the element material $n=1.5$. For these parameters, the optical path length from the plane $z=0$ to the plane $z = f_0 = 10\mm$ is $L = (n-1) h_0 + f_0 =12.5\mm$.

The mapping $\mat{x}(\mat{u})=T(\mat{u})$ was found by solving the LAP of Eq.~\eqref{GrindEQ__8_}. For the calculation of the mapping, the rectangle $D$ was represented as a set of $N = 143^2 = 20449$ rectangular cells $d_i$ with the size $0.0282 \times 0.0141\mm^2$. The circular domain $G$ was also approximated by 20449 square cells $g_i$ with the size $0.0124 \times 0.0124\mm^2$. For the solution of the LAP~\eqref{GrindEQ__8_}, the auction algorithm~\cite{17} implemented in~\cite{18} was used. For the considered example, the time of solving the LAP was less than 3 minutes. For comparison, let us note that the calculation of this mapping using finite-difference methods for solving PDE of Monge--Amp{\`e}re type~\cite{6,7,8} requires the solution of a system of 20449 nonlinear equations and is by several orders of magnitude slower~\cite{16}.

\begin{figure}
	\centering
		\includegraphics{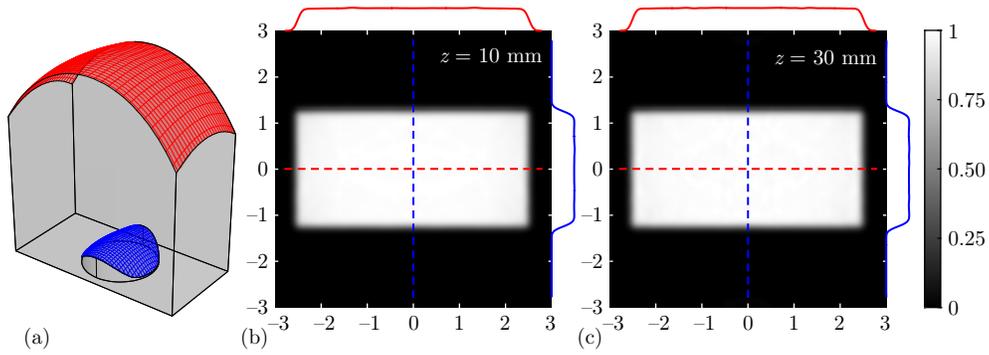}
	\caption{\label{fig2}(a)~Surfaces of an optical element transforming a circular beam into a beam with rectangular cross-section. (b,\,c)~Normalized irradiance distributions generated by the element in the planes $z=10\mm$ and $z=30\mm$ calculated using TracePro. The irradiance cross-sections along the coordinate axes are shown at the top and at the right of the distributions.}
\end{figure}

\begin{figure}
	\centering
		\includegraphics{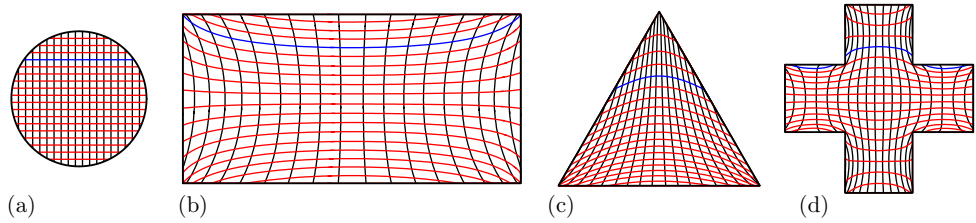}
	\caption{\label{fig3}The mappings $T:G\to D$ corresponding to the transformation of a circular beam~(a) to a rectangular beam~(b), a triangular beam~(c), and a cross-shaped beam~(d). Blue lines mark a straight line in the $G$ domain and its images in the generated domains $D$.}
\end{figure}

The optical surfaces of the element were reconstructed from the computed discrete mapping using the technique described above in Section~\ref{sec4}.
The function $g(\mat{x})$ was represented as a two-dimensional B-spline at $q = 4$ and $N_1 =N_2 =10$. 
The coefficients of the spline were found using the least squares method from the condition of minimizing expression~\eqref{GrindEQ__10_}. 
Then, the second optical surface defined as spline in Eq.~\eqref{GrindEQ__9_} was recalculated on a uniform $400 \times 400$ rectangular grid ($N_g = 400^2$). Then, using Eqs.~\eqref{GrindEQ__12_} and~\eqref{GrindEQ__13_}, the first optical surface was calculated on a uniform $143 \times 143$ square grid. The resulting optical surfaces are shown in Fig.~\ref{fig2}a. For the sake of illustration, Figs.~\ref{fig3}a and \ref{fig3}b depict the mapping $T(\mat{u})$ calculated using Eq.~\eqref{GrindEQ__14_}. Figure~\ref{fig3}a shows a square grid in the domain $G$, whereas Fig.~\ref{fig3}b shows its image by the mapping $T(\mat{u})$. It should be noted that in the considered example, the angles of ray deflection by the optical surfaces (which reach maximum at the corners of the rectangle $D$) are as large as $25\degg$. Hence, the optical element is designed in a non-paraxial case.

In order to verify the proposed design method, the performance of the optical element was numerically simulated using the commercial ray-tracing software TracePro~\cite{27}. 
To do this, the optical surfaces shown in Fig.~\ref{fig2}a were exported to a computer-aided design software Rhinoceros~\cite{28}, in which a 3D-model of the element was created (Fig.~\ref{fig2}a). 
Then this model was exported to TracePro for simulation.
Figures~\ref{fig2}b and~\ref{fig2}c show the simulated normalized irradiance distributions generated by the optical element in the planes $z=10\mm$ and $z=30\mm$. 
The simulation results demonstrate the formation of a rectangle with the required size and near-uniform irradiance. 
The normalized root-mean-square deviations (NRMSD) of the numerically simulated irradiance distributions from a constant value amount to 5.3\% and 6.1\% at $z=10\mm$ and $z=30\mm$, respectively. 
The fact that the size of the rectangle remains unchanged at different distances from the element shows that the wavefront of the beam generated by the element is nearly plane. Actually, the root-mean-square deviation of the optical path length from a constant value at $z = 10\mm$ does not exceed 1~nm.

\subsection{Transformation of a circular beam to a triangular beam}

\begin{figure}
	\centering
		\includegraphics{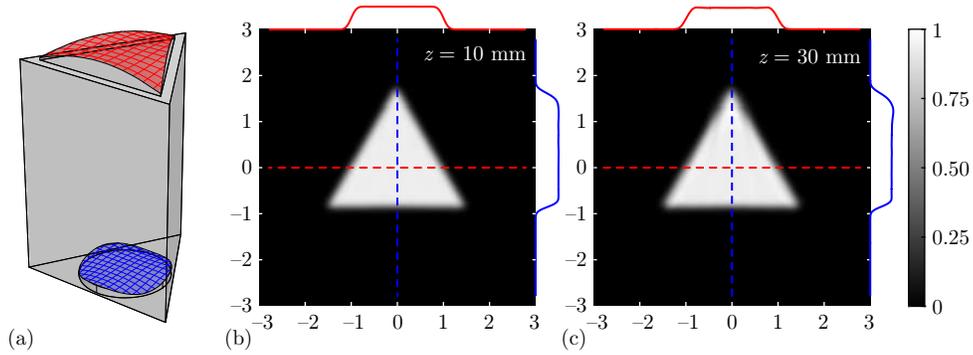}
	\caption{\label{fig4}(a)~Surfaces of an optical element transforming a circular beam to a triangular one. (b,\,c)~Normalized irradiance distributions generated by the element in the planes $z=10\mm$ and $z=30\mm$ calculated using TracePro. The irradiance cross-sections along the coordinate axes are shown at the top and at the right of the distributions.}
\end{figure}

As a second example, let us consider the design of an optical element transforming a uniform circular incident beam to a uniform triangular beam. The equilateral triangle (the domain $D$) has a 3-mm side, with the rest geometric and simulation parameters ($R, f_0, h_0, n, N, q, N_1, N_2, N_g$) remaining the same.
The designed optical surfaces and the mapping $T(\mat{u})$ are depicted in Fig.~\ref{fig4}a and Figs.~\ref{fig3}a and~\ref{fig3}c, respectively.
Figures~\ref{fig4}b and~\ref{fig4}c show the simulated normalized irradiance distributions generated by the optical element in the plane $z=10\mm$ and $z=30\mm$ obtained using TracePro.
The simulation results show that a triangle-shaped irradiance domain with the required size is generated. 
The NRMSD of the calculated irradiance distributions from a uniform value amount to 5.6\% and 6.4\% at $z=10\mm$ and $z=30\mm$, respectively. As in the previous example, the root-mean-square deviation of the optical path length from a constant value at  $z=10\mm$ does not exceed 1~nm.

\subsection{Transformation of a circular beam to a cross-shaped beam}\label{sec53}

\begin{figure}
	\centering
		\includegraphics{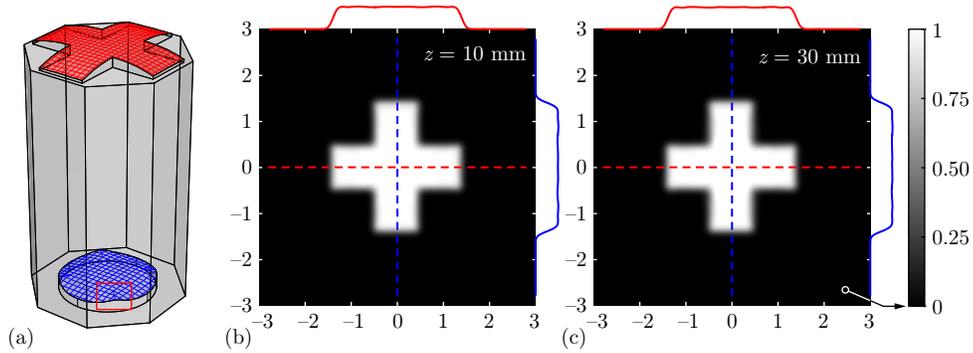}
	\caption{\label{fig5}(a)~Surfaces of an optical element transforming a circular beam to a cross-shaped beam. A magnified fragment of the lower optical surface shown in Fig.~\ref{fig6} is highlighted with a rectangle. (b,\,c)~Normalized irradiance distributions generated by the element in the planes $z=10\mm$ and $z=30\mm$ calculated using TracePro. The irradiance cross-sections along the coordinate axes are shown at the top and at the right of the distributions.}
\end{figure}

\begin{figure}
	\centering
		\includegraphics{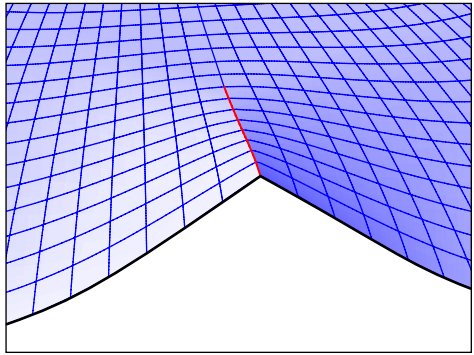}
	\caption{\label{fig6}A magnified fragment of the lower surface [$f(\mat{x})$] of the designed optical element (see Fig.~\ref{fig5}a). The break is marked with a red line.}
\end{figure}

The final and the most challenging example involves the design of an optical element transforming a uniform incident beam with circular cross-section into a beam with cross-shaped uniform irradiance distribution. In this case, the domain $D$ corresponds to a cross obtained by a superposition of two mutually perpendicular rectangles with the sizes $2.8\times1\mm^2$ and $1\times2.8\mm^2$. The rest geometric and calculation parameters remain the same. The resulting optical surfaces are shown in Fig.~\ref{fig5}a. Note that the lower surface of the optical element has `breaks' (sharp bends) along the bisector in each quadrant of the coordinate system. Figure~\ref{fig6} depicts a magnified fragment of the lower optical surface with a break, which is highlighted by a red rectangle in Fig.~\ref{fig5}a. 
These breaks are caused by the discontinuities in the mapping $\mat{x}(\mat{u})=T(\mat{u})$. 
Indeed, the rays that strike the lower optical surface on the opposite sides of the bisector are mapped to different parts of the cross. 
This is confirmed by the form of the mapping $T(\mat{u})$ shown in Fig.~\ref{fig3}d. 
For instance, the image of a straight blue line shown in Fig.~\ref{fig3}a has discontinuities near the inner corners of the cross and is composed of several curvilinear segments. 
It is worth noting that since the inverse mapping $T^{-1}: D\to G$ is continuous, the obtained upper optical surface is smooth.

The simulation results obtained using TracePro and shown in Figs.~\ref{fig5}b and~\ref{fig5}c demonstrate that a high-quality cross-shaped irradiance distribution is generated. The NRMSD of the calculated irradiance distributions from a constant value are 7.1\% and 7.9\% at $z=10\mm$ and $z=30\mm$, respectively. The root-mean-square deviation of the optical path length from a constant value is 1.1~nm at $z = 10\mm$. The authors believe that the somewhat larger NRMSD of the generated irradiance distributions from a constant value is due to the approximation error of the piecewise-smooth lower optical surface in the used CAD software Rhinoceros. This software adopts non-uniform rational basis splines (NURBS) for representing a surface from a point cloud. As a result, the sharp bends of the lower optical surface are somewhat smoothed.

The third example demonstrates one of the advantages of the proposed method, which is the capability of operating with discontinuous mappings. In this case, the known methods based on the numerical solution of an elliptic PDE~\cite{6,7,8,9,10,11} are either not applicable or numerically unstable.

\section{Conclusion}

We proposed a method for designing refractive optical elements for collimated beam shaping. The computation of a ray mapping is reduced to a mass transportation problem with a non-quadratic cost function. In the proposed method, the problem of computing a ray mapping is formulated as a linear assignment problem, which constitutes a discrete version of the corresponding mass transportation problem. A technique for reconstructing an optical surface from the computed ray mapping has also been proposed. As distinct from the methods based on the numerical solution of an elliptic PDE~\cite{6,7,8,9,10,11}, the method proposed here is suitable for the design of optical elements with continuous piecewise-smooth optical surfaces.

We have designed refractive optical elements transforming a beam with circular cross-section into variously shaped (rectangular, triangular, and cross-shaped) beams with plane wavefronts. The proposed method is computationally efficient. It requires less than 3~minutes on a standard computer to compute a ray mapping on a $143 \times 143$ grid. High performance of the designed optical elements is confirmed by the numerical simulation results.

\section*{Funding}
Ministry of Education and Science of Russian Federation; 
Russian Foundation for Basic Research (RFBR) grant **-**-****; 
*****.

\end{document}